# On some peculiarities of electric field pulse propagation in electron Maxwellian plasma and its back response*

V. N. Soshnikov†

Plasma Physics Dept.,
All-Russian Institute of Scientific and Technical Information
of the Russian Academy of Sciences
*(VINITI, Usievitcha 20, 125315 Moscow, Russia)*

**Abstract**

*In the spirit of continued study of plasma wave general properties we investigated the boundary problem with the simplest form of electric field pulse $E_0 \exp(-\alpha t)$ at the edge $x = 0$ of half-infinite uniform plasma slab. In the case of longitudinal electric field pulse its traveling velocity is essentially other than in the case of harmonic waves; there is also no back response. In the case of transverse field pulse there takes place the bimodal propagation rate of the fast pulse signal with amplitude $E_0$ and weak slow sign reversed pulse signals; some very weak response ("echo") arises with a time delay of the order $\dfrac{c}{v_{eff}} \times \dfrac{2}{\alpha \sqrt{1 + \omega_L^2/\alpha^2}}$ in the near coordinate zone of formation of the asymptotical regime, where $v_{eff}$ is some characteristic velocity $\sim (1.5 \div 2)\sqrt{k_B T / m_e}$ for Maxwellian distribution; $c$ is light velocity, $\omega_L$ is Langmuir electron frequency.*

**Introduction**

   In the preceding papers (see [1] and references therein) we have considered some general features of 1D electron waves in a half-infinite homogeneous plasma slab which are excited by harmonic electric field at a boundary plane of the slab. These investigations were based on a simple and appearing rather evident principle: indefinitely divergent integrals (IDI) emerging in the dispersion equations are a direct consequence of the information lack in original wave equations (kinetic and Maxwell equations). Thus method of calculation for IDI has to be defined neither arbitrarily nor from ubiquitous mathematical analogies, but strongly follows from some additional physical conditions of the problem resulting in its single solution. It is significant here that all diverse asymptotical solutions obtained with the Laplace transform method are satisfying original wave equations independently on understanding (on implied sense) of the IDI.

   For the boundary problem with waves excited by harmonic electrical field such physical conditions are both impossibility of kinematical waves which would not be bound with the boundary electric field, and absence of backward traveling waves, at least for the fast wave mode, since in the homogeneous plasma slab there are no sources of exciting them. It is shown that these complementary conditions are satisfied for IDI in the principal value sense (Vlasov) but are not satisfied with IDI in the contour integral sense with passing around poles in the complex plane of integration in $v_x$ (Landau). It ought also to note that these additional requirements define the missing boundary condition for the electron distribution function in the only way.

   Before used Laplace transform method for obtaining asymptotical solution can now be used with some reservation also in the case of pulse boundary field. Since we are interested in the most general in principle difference of the pulse and harmonic excitations, we have selected $E_0 \exp(-\alpha t)$ as a simple illustrative boundary pulse excitation field with its simplest Laplace transform

$$g(p_1) = E_0 \int e^{-\alpha t - p_1 t} dt = \frac{E_0}{\alpha + p_1} \equiv E_{p_1}, \qquad (1)$$

where $p_1$ is Laplace transform parameter in time; in the following $p_2$ is transform parameter in coordinate $x$ along which the perturbation is propagating. One supposes Maxwellian plasma for which calculation of the principal value of IDI is drastically facilitated due to replacement of the integrating value $v_x$ with some characteristic value

---

$$v_{eff} \sim (1.5 \div 2)\sqrt{\overline{v_x^2}} \qquad (2)$$

(see [1], [2], [3] ).

## Longitudinal field pulse

We use expressions for Laplace images in [4]

$$E_{p_1 p_2} = \frac{1}{p_2 G(p_1 p_2)} \left( E_{p_1 p_2} + \int \frac{v_x f_{p_1}}{p_1 + v_x p_2} d\vec{v} \right), \qquad (3)$$

$$G(p_1 p_2) = 1 + \frac{\omega_L^2}{p_2} \int \frac{\partial f_0 / \partial v_x}{p_1 + v_x p_2} d\vec{v} \approx 1 + \frac{\omega_L^2}{p_1^2 - v_{eff}^2 p_2^2}, \qquad (4)$$

and taking into account that $E_{p_1}$, as well as $f_{p_1}$ have pole $p_1^{(0)} = -\alpha$, in analogy with the harmonic boundary perturbation, one obtains an equation for determination the pole $p_2^{(0)}$

$$1 + \frac{\omega_L^2}{\alpha^2 - v_{eff}^2 p_2^{(0)2}} = 0, \qquad (5)$$

hence

$$p_2^{(0)} = \pm \sqrt{\frac{\alpha^2 + \omega_L^2}{v_{eff}^2}}. \qquad (6)$$

Correspondingly, for the electric field image $E_{p_1 p_2}$ original is

$$E(x,t) = E_1 \exp\left( -\alpha t \pm \frac{\alpha}{v_{eff}} \sqrt{1 + \omega_L^2 / \alpha^2} \, x \right), \qquad (7)$$

where $E_1$ may be different from $E_0$ owing to Eq. (3) and the additional boundary conditions which determine the form of a boundary perturbation $f_1$ [1], [4].

In the round brackets of Eq. (7) it ought to select the sign minus which corresponds to a forward pulse since there are no physical sources of a backward pulse in homogeneous plasma as it was discussed in preceding papers [1], [4] and was explained presumably by forming some suited boundary polarization. But then the pulse propagates only in the forward direction, and Eq. (7) has the physical sense at

$$0 \leq v_{eff} t - \sqrt{1 + \frac{\omega_L^2}{\alpha^2}} x < \infty. \qquad (8)$$

This circumstance might be related with some purely mathematical constraints on Laplace transform. So, pulse velocity is now

$$v_{pulse} = \frac{v_{eff}}{\sqrt{1 + \omega_L^2 / \alpha^2}}. \qquad (9)$$

It appears to be sound an assumption that pulse formation is lasting within the characteristic boundary excitation time $t_{excit} \sim (1 \div 2)/\alpha$. It is thus a characteristic time of setting asymptotical solution. In this case there are no reasons to expect appearance of a back signal of the longitudinal field pulse ("echo"). This problem can be solved both experimentally and theoretically, for example with numerical solution of the coupled original equations at boundary conditions for the field and distribution function. These conditions have to be found before from Laplace transform with additional asymptotic conditions of absence the back pulse and absence of kinematical modes which would not bound with the given boundary electric field.

## Transverse field pulse

In the case of transverse field pulse one can use the solutions for transverse waves in [5] replacing there $p_1^{(0)} = i\omega$

with $p_1^{(0)} = -\alpha$ and interchanging forward waves with moving backward pulse, and backward wave with the traveling forward pulse.

Characteristic equation in [4] for $p_2^{(0)}$ in the case of pulse with $p_1^{(0)} = -\alpha$ will be

$$G(-\alpha, p_2) = p_2^2 - \frac{\alpha^2}{c^2} + \frac{\omega_L^2 \alpha}{c^2} \int v_z \frac{\partial f_0 / \partial v_z}{\alpha - v_x p_2} d\vec{v} \approx p_2^2 - \frac{\alpha^2}{c^2} \left(1 + \frac{\omega_L^2}{\alpha^2 - v_{eff}^2 p_2^2}\right) = 0 \qquad (10)$$

with the solution

$$p_2^{(0)} = \pm \sqrt{\frac{\alpha^2 (c^2 + v_{eff}^2) \pm \sqrt{\alpha^4 (c^2 + v_{eff}^2)^2 - 4\alpha^2 \omega_L^2 c^2 v_{eff}^2}}{2 c^2 v_{eff}^2}}, \qquad (11)$$

and with the physical condition of absence of any sources which could increase pulse energy,

$$0 \leq \alpha t - p_2^{(0)} x < \infty. \qquad (12)$$

Accounting for the smallness of $v_{eff}^2 / c^2 \ll 1$ one obtains

$$\sqrt{\alpha^4 (c^2 + v_{eff}^2)^2 - 4\alpha^2 \omega_L^2 c^2 v_{eff}^2} \approx \alpha^2 (c^2 + v_{eff}^2) \left(1 - 2\frac{\omega_L^2}{\alpha^2} \frac{c^2 v_{eff}^2}{(c^2 + v_{eff}^2)^2}\right). \qquad (13)$$

Corresponding roots of the dispersion equation are then

$$p_2^{(1\pm)} \approx \pm \frac{\alpha}{c} \left(1 + \frac{\omega_L^2}{\alpha^2}\right)^{1/2}, \qquad (14)$$

$$p_2^{(2\pm)} \approx \pm \frac{\alpha}{v_{eff}} \left(1 - \frac{1}{2} \frac{\omega_L^2}{\alpha^2} \frac{v_{eff}^2}{c^2}\right). \qquad (15)$$

The root $p_2^{(1+)}$ relates to the pulse velocity

$$v_{pulse}^{(1)} \approx \frac{c}{\sqrt{1 + \omega_L^2 / \alpha^2}}. \qquad (16)$$

The root $p_2^{(2\pm)}$ relates to the pulse velocity

$$v_{pulse}^{(2)} \approx \pm \frac{v_{eff}}{1 - \frac{1}{2} \frac{\omega_L^2}{\alpha^2} \frac{v_{eff}^2}{c^2}} \approx \pm v_{eff}. \qquad (17)$$

However, as it was noted before, there are no reflection sources in uniform plasma for the rapid pulse mode, so at $p_1^{(0)} = -\alpha$ it ought to pick out only the root $p_2^{(1+)}$ for the high velocity pulse. So far as the asymptotic solution for the rapid pulse is forming still during some characteristic time of exciting exponential pulse $t_{excit} \sim 2/\alpha$, the rapid pulse is running a distance

$$x_{rap} \sim t_{excit} v_{pulse}^{(1)} = \frac{2}{\alpha} \frac{c}{\sqrt{1 + \omega_L^2 / \alpha^2}}. \qquad (18)$$

Within this time both the forward and backward pulses being formed, the last one being returned to the boundary plane ("echo" effect) within a time

$$t_{return} \sim \frac{2c}{\alpha v_{eff} \sqrt{1+\omega_L^2/\alpha^2}} \quad . \qquad (19)$$

**Amplitudes of transverse field pulses**

Amplitudes of forward and backward pulses can be evaluated in all analogy with the evaluation of amplitudes of the slow and rapid transverse waves in [5]. A condition of absence of the backward rapid pulse is

$$p_2^{(1-)} E_0 + F_{p_1} = 0 \quad , \qquad (20)$$

where $F_{p_1}$ is Laplacian image of the boundary function $\left.\frac{\partial E}{\partial x}\right|_{x=0}$, hence we have

$$F_{-\alpha} = \frac{\alpha}{c} E_0 \sqrt{1+\omega_L^2/\alpha^2} \quad . \qquad (21)$$

Taking into account that characteristic equation for $p_2$ can be written as

$$G(-\alpha, p_2) = A(-\alpha, p_2)\left(p_2 - p_2^{(1+)}\right)\left(p_2 - p_2^{(1-)}\right)\left(p_2 - p_2^{(2+)}\right)\left(p_2 - p_2^{(2-)}\right) = 0 \qquad (22)$$

where

$$A(-\alpha, p_2) = \frac{v_{eff}^2}{-\alpha^2 + v_{eff}^2 p_2^2} \quad , \qquad (23)$$

one obtains

$$\left(\frac{RES}{2\pi i}\right) E_{-\alpha, p_2^{(1+)}} = \left(\frac{RES}{2\pi i}\right) \frac{p_2^{(1+)} E_0 + F_{-\alpha}}{G(-\alpha, p_2^{(1+)})} \approx E_0 \qquad (24)$$

with following calculation of residuum in the case of rapid forward pulse (that is calculating its amplitude) neglecting small terms of the order $v_{eff}^2/c^2$.

Analogously for the amplitude of the forward slow pulse one obtains

$$\left(\frac{RES}{2\pi i}\right) E_{-\alpha, p_2^{(2+)}} = \left(\frac{RES}{2\pi i}\right) \frac{p_2^{(2+)} E_0 + F_{-\alpha}}{G(-\alpha, p_2^{(2+)})} = -\frac{E_0}{2} \frac{v_{eff}^2}{c^2} \frac{\omega_L^2}{\alpha^2} \quad , \qquad (25)$$

and for the amplitude of the backward slow pulse one obtains

$$\left(\frac{RES}{2\pi i}\right) E_{-\alpha, p_2^{(2-)}} = \left(\frac{RES}{2\pi i}\right) \frac{p_2^{(2-)} E_0 + F_{-\alpha}}{G(-\alpha, p_2^{(2-)})} = -\frac{E_0}{2} \frac{v_{eff}^2}{c^2} \frac{\omega_L^2}{\alpha^2} \quad . \qquad (26)$$

In this way amplitudes of the forward and the backward slow pulses are much smaller than amplitude of the rapid forward pulse, and both foregoing ones have reversed sign relative to the rapid pulse.

Emerging of returning signal is perceived as a reflection of the forward slow mode pulse of the solution, however it appears more sound to take this effect as a result of the polarization charge redistribution in the near zone of forming asymptotic solution for the rapid pulse, such that the rapid backward pulse would be prevented.

It ought to note that the presented considerations confirm the possibility to satisfy additional boundary conditions and to determine the form of distribution function and $\partial E/\partial x$ at the boundary just at the IDI taken in the principal value sense.

**Conclusion**

We have considered propagation of pulses in a half-infinite slab of homogeneous Maxwellian electron plasma which were excited by boundary pulses of longitudinal or transverse electric field, using some rather simplified illustration of Laplace transform method. It ought to note the sign reversing of the transverse slow mode amplitude relative to the rapid one and significant dependence of the pulse velocity on its form.

We used a robust assumption that setting of the asymptotical solution is occurring at the early stage in the near boundary polarization zone within the rapid pulse length where some polarization charge redistribution occurs. In this case, for the pulse excited by longitudinal boundary field the backward pulse does not arise. Its arising ("echo") is possible only in the case of transversal exciting field at the bimodal solution. The rapid pulse goes ahead, it is followed with some lag by the slow pulse. In the near boundary zone of forming the rapid pulse the backward slow pulse is also forming with a response time lag, the last one is defined with the length of the rapid pulse and the backward slow pulse velocity.

It is significant that this considered although simple however specific case of electric signal backward response is related to neither additional non-linear terms in original linear equations, nor Landau damping and resonance electron beams, and it does not pretend to any generalization of a huge variety of echo phenomena in very divers methods of supporting and influence upon plasmas (see, for example, [6] ).